\documentclass[twocolumn,showpacs,nofootinbib]{revtex4}
\usepackage{graphicx}
\usepackage{amsmath,amssymb}

\begin{document}

\title{The variation of the electromagnetic coupling and quintessence}

\author{M. C. Bento}
\email{bento@sirius.ist.utl.pt}
\affiliation{Centro de F\'{\i}sica Te\'orica de
Part\'{\i}culas, Instituto Superior T\'{e}cnico\\ Avenida Rovisco
Pais, 1049-001 Lisboa, Portugal}

\author{R. Gonz\'{a}lez Felipe}
\email{gonzalez@cftp.ist.utl.pt}
\affiliation{%
Instituto Superior de Engenharia de Lisboa\\ Rua Conselheiro Em\'idio Navarro, 1959-007 Lisboa,
Portugal}
\affiliation{%
Centro de F\'{\i}sica Te\'orica de
Part\'{\i}culas, Instituto Superior T\'{e}cnico\\ Avenida Rovisco
Pais, 1049-001 Lisboa, Portugal}

\date{\today}

\begin{abstract}

The properties of quintessence are examined through the study of the variation of the electromagnetic coupling. We consider two simple quintessence models with a modified exponential potential and study the parameter space constraints derived from the existing observational bounds on the variation of the fine structure constant and the most recent Wilkinson Microwave Anisotropy Probe observations.

\vskip 0.5cm

\end{abstract}

 \pacs{98.80.-k,98.80.Cq,12.60.-i}

\maketitle
\vskip 2pc
\section{Introduction}

Over the last decade, the temporal and spatial variation of fundamental constants has become a very popular subject. The motivation partially comes from theories unifying gravity and other interactions, which suggest that fundamental constants could have indeed varied during the evolution of the universe~\cite{fund-th}. It is therefore particularly relevant to search for these variations and try to establish correlations, if any, with other striking properties of the universe, as for instance with dark energy.

From the observational point of view, the time variation of the fine structure constant $\alpha$ has been widely discussed in several contexts. In particular, from the spectra of quasars (QSO), one obtains from the KecK/HIRES instrument~\cite{quasars1}
\begin{align}
\frac{\Delta\alpha}{\alpha}=(-0.57 \pm 0.11)\times 10^{-5}~,~\text{for}~ 0.2 < z < 4.2~,
\label{murphy}
\end{align}
while the Ultraviolet and Visual Echelle Spectrograph (UVES) instrument~\cite{quasars2, quasars3}
implies
\begin{align}
\frac{\Delta\alpha}{\alpha}=(-0.64 \pm 0.36)\times 10^{-5}~,~\text{for} ~0.4< z < 2.3~.
\label{chand}
\end{align}
The Oklo natural reactor also provides a bound,
\begin{align}
-0.9\times 10^{-7}<\frac{\Delta\alpha}{\alpha}<1.2\times 10^{-7}~,~\text{for}~z<0.14~,
\label{oklo}
\end{align}
at $95 \%$ C.L.~\cite{oklo1,oklo2,oklo3}. Furthermore, estimates of the age of iron meteorites, corresponding to $z \simeq 0.45$, when combined with the measurement of the Os/Re ratio resulting from the radioactive decay $^{187}$Re $\to\, ^{187}$Os, yield \cite{met1,met2,met3}:
\begin{align}
\frac{\Delta\alpha}{\alpha}=(-8 \pm 8) \times 10^{-7}~,
\label{meteorites1}
\end{align}
at $1\sigma$, and
\begin{align}
-24\times 10^{-7}<\frac{\Delta\alpha}{\alpha}<8\times 10^{-7}~,
\label{meteorites2}
\end{align}
at $2\sigma$.

Big bang nucleosynthesis (BBN) also places bounds on the variation of $\alpha$~\cite{BBN}:
\begin{align} \label{BBN}
-0.05<\frac{\Delta\alpha}{\alpha}<0.01~,~\text{for}~10^9<z<10^{10}~.
\end{align}
Finally, the 5-year data from the Wilkinson Microwave Anisotropy Probe (WMAP) with Hubble Space Telescope (HST) prior provides the following bound at 95\% C.L.~\cite{Nakashima:2008cb}:
\begin{align} \label{WMAP}
-0.028 <\frac{\Delta\alpha}{\alpha}< 0.026~, ~\text{for}~z \sim 10^3.
\end{align}
Without HST prior this bound is less restrictive~\cite{Nakashima:2008cb}:
\begin{align} \label{WMAP2}
-0.050 <\frac{\Delta\alpha}{\alpha}< 0.042~,
\end{align}
at 95\% C.L.

On the other hand, recent observations of high redshift type Ia supernova and, more indirectly, of the CMB and galaxy clustering, indicate that the universe is undergoing a period of accelerated expansion~\cite{expansion}. This suggests that the universe is dominated by a form of energy density with negative pressure (dark energy). An obvious candidate for dark energy could be an uncanceled cosmological constant~\cite{BentoUncanceledCC}, which however would require an extremely high fine-tuning. Quintessence-type models~\cite{quint} with one~\cite{quint1} or two~\cite{quint2} scalar fields, k-essence~\cite{k-ess} and the Chaplygin gas with an exotic equation of state~\cite{GCG3,bil} are among other possibilities.

In most of the theoretical approaches, scalar fields are present in the theory. Thus, one could expect that the two observational facts, namely the variation of $\alpha$ and the recent acceleration of the universe, are somehow related. Indeed, the coupling of such a scalar field to electromagnetism would lead to a variation of the fine structure constant~\cite{Bekenstein}. In several contexts~\cite{carroll,supergrav,mohammad,BM}, the above question has already been addressed. 

In this work we shall consider two simple quintessence models with a modified exponential potential, proposed by Albrecht and Skordis~\cite{skordis1,skordis2}. Our aim is to restrict the parameter space of these models by using the observational constraints on the variation of the fine structure constant and by imposing consistency with the 5-year data from WMAP.


\section{The models}

\subsection{Quintessence}

The authors of Refs.~\cite{skordis1, skordis2} add a polynomial prefactor in front of the exponential potential in order to introduce a local minimum in the potential $V(\phi)$, such that the scalar field $\phi$ gets trapped into it. The potential can be written in the form
\begin{align}
\label{potAS1}
V(\phi) = V_0 [(\phi - \phi_0)^2 + A]\, e^{-\lambda \phi}.
\end{align}
Since the effect of trapping is equivalent to a cosmological constant, an era of accelerated expansion is eventually reached.  Notice that this model has already been analyzed in the context of the variation of the fine structure constant~\cite{anchordoqui}. However, the authors consider a particular set of values for the parameters in the potential, whereas here we vary these parameters and try to find constraints on them. Recently, this potential has also been investigated in the braneworld context as a model of quintessential inflation~\cite{Bento:2008yx}. We shall refer to this potential as the AS1 model.

Regardless of initial conditions, trapping occurs in the very early universe, when the field enters the attractor regime. The tracking properties of the AS1 potential~(\ref{potAS1}) are similar to those of a pure exponential potential. Nevertheless, due to the presence of the polynomial factor, quintessence dominates near the present epoch. Furthermore, this potential can lead to both permanent and transient acceleration regimes. When $A\lambda^2 < 1$ and the field is trapped in the local minimum of the potential, permanent acceleration is achieved. On the other hand, if $A\lambda^2 < 1$ and the field arrives at the minimum with a kinetic energy sufficiently high to roll over the barrier, a transient acceleration regime is obtained~\cite{Barrow:2000nc}. Transient vacuum domination also arises for $A\lambda^2 >1$, when the potential loses its local minimum~\cite{Barrow:2000nc}. The existence of a transient regime is interesting from the theoretical viewpoint of string theory, since it can avoid the difficulties which typically arise in the S-matrix construction at the asymptotic future in a de-Sitter space~\cite{Hellerman:2001yi, Fischler:2001yj, Witten:2001kn}.

A second potential (hereafter referred as AS2 model) which can lead to the desired accelerated expansion has the form~\cite{skordis2},
\begin{align}
 \label{potAS2}
V(\phi) = \left[ {C\over (\phi-\phi_0)^2+A} +D\right] e^{-\lambda \phi}~.
\end{align}
The motivation of such a potential comes from brane studies, which indicate that it could arise as a Yukawa-like interaction between branes~\cite{Dvali:1998pa}. The behavior of the AS2 potential is similar to the AS1 potential during the radiation-dominated era. Nevertheless, during matter domination it retains much more density than the AS1 potential. This is due to the fact that the AS2 potential has a smoother minimum and a sharper maximum when compared to the AS1 one. Furthermore, acceleration is achieved earlier for the latter potential~\cite{skordis2}.

The evolution of the scalar field is described by the equation of motion
\begin{align}
{\ddot \phi} + 3H {\dot \phi} + V'(\phi) = 0~,
\label{eq:kg}
\end{align}
where the dot and prime denote derivatives with respect to time and $\phi$, respectively;
\begin{align}
H^2 = {1 \over 3}\, \rho\,,\quad  \rho=\rho_r + \rho_m + \rho_\phi~.
\label{eq:Friedmann}
\end{align}
Here $\rho_{r}$ and $\rho_{m}$ are the radiation and matter energy densities, respectively, and $\rho_\phi = \dot\phi^2/2 + V(\phi)$. In order to integrate equations (\ref{eq:kg}) and (\ref{eq:Friedmann}), it is more convenient to rewrite them in the form
\begin{align}
{d x \over d \cal{N}} & = -3 x + \sigma \sqrt{3\over 2}\,y^2+{3\over 2}\, x \,[2 x^2+\gamma(1-x^2-y^2)]~,\nonumber\\
{d y \over d \cal{N}}  & = -\sigma\sqrt{3\over 2}\, x\, y + {3\over 2}\, y\, [2 x^2+\gamma(1-x^2-y^2)],
\label{eq:system}
\end{align}
where $\gamma = w_B+1$, $w_B$ is the equation-of-state parameter for the background ($w_B=0, 1/3$ for matter and radiation, respectively); ${\cal N} \equiv \ln\, a$, with $a$ being the scale factor, and
\begin{align}
x \equiv {\dot{\phi} \over \sqrt{2\,\rho}}~,\quad
y \equiv   {\sqrt{V} \over \sqrt{\rho}}~,\quad
\sigma \equiv -  {V^\prime\over V} ~.
\label{eq:defs}
\end{align}

The above equations are integrated from the Planck epoch ($a \simeq 10^{-30}$) to the present epoch ($a \equiv a_0 =1$). For the fraction of radiation at present we use the central value $\Omega_r^0 \,h^2 = 4.3 \times 10^{-5}$, which assumes the addition of three neutrino species. The matter content is such that the correct fraction is reproduced at present, $\Omega_m^0 \,h^2 = 0.1369\pm 0.0037\,$~\cite{Komatsu:2008hk}. Furthermore, we assume that early in the radiation era the scalar field is in its tracking regime, in which case~\cite{Ferreira:1997hj}
\begin{align}
\rho_\phi\simeq 3V,\quad \Omega_\phi \simeq {4 \over \lambda^2}~.
\end{align}


\subsection{Coupling with electromagnetism}

Due to the universality of the gravitational interactions, non-renormalizable couplings of the quintessence field $\phi$ to the standard model fields are expected below the Planck scale. Following Bekenstein's proposal~\cite{Bekenstein}, one can consider the interaction between the scalar and electromagnetic fields in the form
\begin{align}
{\cal L}_{em} =\frac{1}{16\pi} f(\phi)\,F_{\mu\nu}F^{\mu\nu},
\end{align}
where $f$ is an arbitrary function. Since the variation of the electromagnetic coupling is small, one can expand this function up to first order in powers of $\phi$ and write
\begin{align}
f(\phi)=\frac{1}{\alpha_{0}}\left[1+\xi \left(\phi-\phi_0\right)\right],
\end{align}
where $\xi$ is a constant. It follows then that the fine structure parameter $\alpha$ is given by
\begin{align}
\alpha =\left[f\left(\phi\right)\right]^{-1}=\alpha_{0}\left[1-\xi\left(\phi-\phi_0\right)\right], \label{alpha2}
\end{align}
and its variation is
\begin{align}
\frac{\Delta\alpha}{\alpha}=\xi\left(\phi-\phi_0\right).
\label{delta2}
\end{align}
Moreover, the rate of variation of $\alpha$ at present is given by
\begin{align}
\frac{\dot \alpha}{\alpha}= \xi \frac{d\phi}{d\cal{N}} H_0,
\end{align}
where $H_0=100\, h$~km~s$^{-1}$~Mpc$^{-1}$ is the Hubble constant.

At this point, it is worth noticing that any variation of $\alpha$ should comply with measurements on the violation of the equivalence principle, which translate into the following upper bound on the coupling between quintessence and the electromagnetic field~\cite{equivalence}:
\begin{align}
\xi \leq 7 \times 10^{-4}.
\label{zetaF}
\end{align}

\section{Parameter space constraints}

In our study, we perform a random analysis on the parameters of the quintessence potentials described by Eqs.~(\ref{potAS1}) and (\ref{potAS2}). For the AS1 model we consider the two possible late time behaviors: permanent or transient acceleration. A viable model should comply with several observational bounds during the different stages in the evolution of the universe. A stringent bound comes from the amount of dark energy during nucleosynthesis, $\Omega_\phi^{BBN}(z \approx  10^{10}) \lesssim 0.09$~\cite{BBNCMB}, coming from the primordial abundances of $^4$He and D, which sets the lower bound $\lambda \gtrsim  6.7$. We scan the full parameter space imposing the following conservative bounds at the present epoch:
\begin{align}
0.6 \leq h &\leq 0.8 ,~0.6 \leq \Omega_\phi^0 \leq 0.8~,\nonumber \\
w_\phi^0 &\leq -0.8 ,~ q_0 < 0 ,
\end{align}
where $q \equiv - \ddot{a}/(a H^2)$ is the deceleration parameter.

\begin{figure}[t]
\includegraphics[width=8cm]{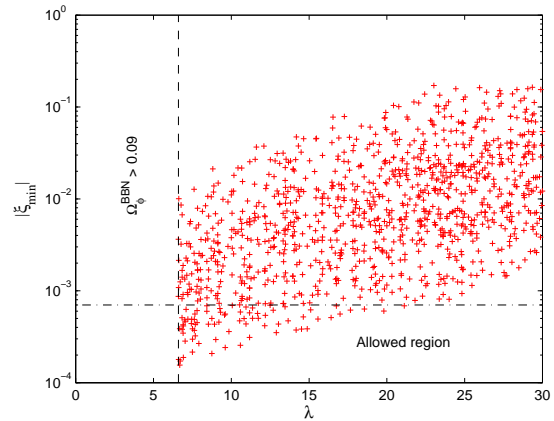}
\caption{The minimum value, $\xi_{min}$, allowed for the coupling $\xi$ between quintessence and the electromagnetic field as a function of $\lambda$, for the AS1 model in the permanent acceleration case. The horizontal dot-dashed line corresponds to the equivalence principle bound (\ref{zetaF}).}
\label{fig:AS1perm1}
\end{figure}
\begin{figure}[t]
\includegraphics[width=8cm]{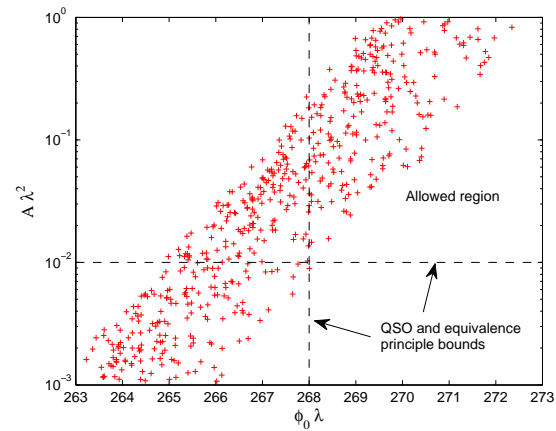}
\caption{Parameter space consistent with all the observational bounds, for the AS1 model in the permanent acceleration case. The horizontal and vertical lines correspond to the lower bounds on $A \lambda^2$ and $\phi_0 \lambda$, respectively, coming from the QSO bounds on the variation of $\alpha$ and the requirement $|\xi_{min}| < 7 \times 10^{-4}$.}
 \label{fig:AS1perm2}
\end{figure}

\begin{figure}[t]
\includegraphics[width=8cm]{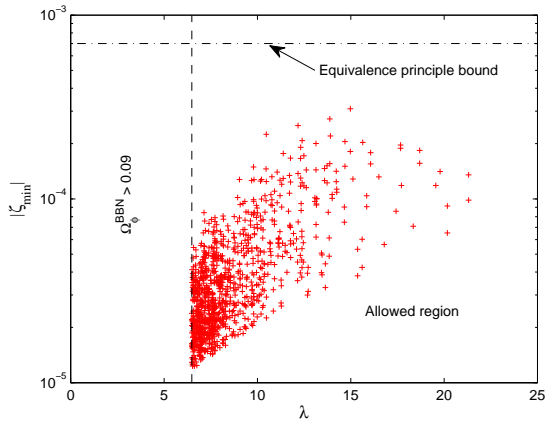}
\caption{As for Figure~\ref{fig:AS1perm1}, but for the transient acceleration case.} \label{fig:AS1transient1}
\end{figure}
\begin{figure}[t]
\includegraphics[width=8cm]{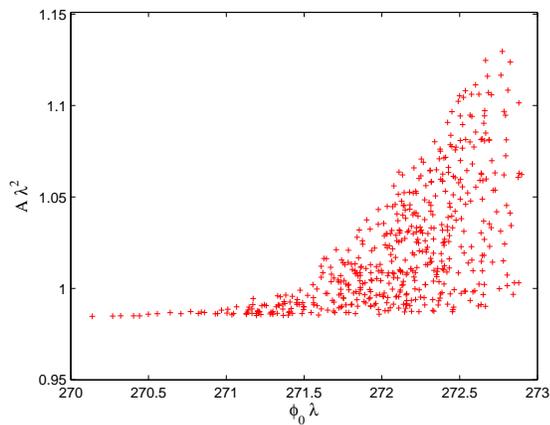}
\caption{As for Figure~\ref{fig:AS1perm2}, but for the transient acceleration case.}  \label{fig:AS1transient2}
\end{figure}

Our results for the AS1 model are presented in Figures \ref{fig:AS1perm1} to \ref{fig:AS1transient2}. The parameter space consistent with all the observational bounds is shown in Figs.~\ref{fig:AS1perm1} and \ref{fig:AS1perm2} for the permanent acceleration case. We remark that the variation of the fine structure constant $\alpha$ coming from the QSO bounds [cf. Eqs.~(\ref{murphy}) and (\ref{chand})] necessarily requires a non-vanishing value of the coupling parameter $\xi$. This imposes a lower bound on this parameter, $\xi_{min}$, for the model to be consistent with the observational data. Clearly, in order to have a consistent solution, the value of $|\xi_{min}|$ should always be lower than the upper bound established by the equivalence principle constraint of Eq.~(\ref{zetaF}). From the figures we conclude that the parameter space of the AS1 model with permanent acceleration is restricted to the ranges
\begin{align}\label{eq:boundperm1}
6.6 &\lesssim \,\lambda\, \lesssim 21.1\,,\nonumber\\
0.01 &\lesssim A \lambda^2 \lesssim 0.97\,,\\
268.1 &\lesssim\, \phi_0 \lambda \lesssim 272.3\,.\nonumber
\end{align}

We recall that for the AS1 potential the combination of parameters $A \lambda^2$ is useful in distinguishing between the permanent and transient regimes, since it determines the presence or absence of the minimum in the potential. Furthermore, the combination  $\phi_0 \lambda$ determines the position of the minimum/maximum or inflection point of the potential, $\phi_\pm = (1+ \phi_0 \lambda \pm \sqrt{1-A \lambda^2})/\lambda$, which in turn is related to the exit from the tracking regime and to the scalar field energy density domination at present.

The results for the transient acceleration case are presented in Figs.~\ref{fig:AS1transient1} and \ref{fig:AS1transient2}. In this case, the equivalence principle and QSO bounds on the variation of the electromagnetic coupling do not yield an upper bound on the parameter $\lambda$ more restrictive than the one already imposed by other observational constraints (see Fig.~\ref{fig:AS1transient1}). We find the following allowed ranges for the parameters of the AS1 model with transient acceleration:
\begin{align}\label{eq:boundtrans1}
6.5 &\lesssim \,\lambda\, \lesssim 21.3\,,\nonumber\\
0.98 &\lesssim A \lambda^2 \lesssim 1.13\,,\\
270.1 &\lesssim\, \phi_0 \lambda \lesssim 272.9\,.\nonumber
\end{align}

\begin{figure}[ht]
\includegraphics[width=8cm]{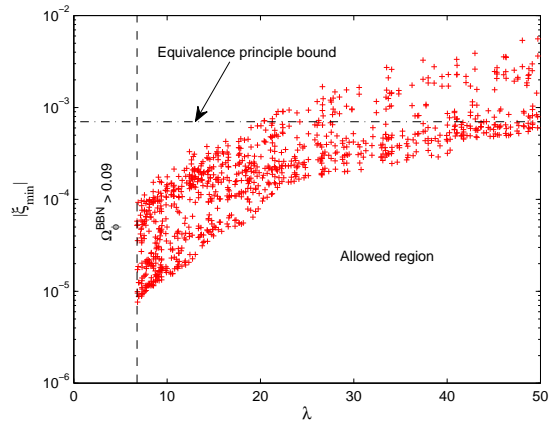}
\caption{$\zeta_{min}$ as a function of $\lambda$ for the AS2 model.}
\label{fig:AS21}
\end{figure}

\begin{figure*}[t]
\includegraphics[width=16cm]{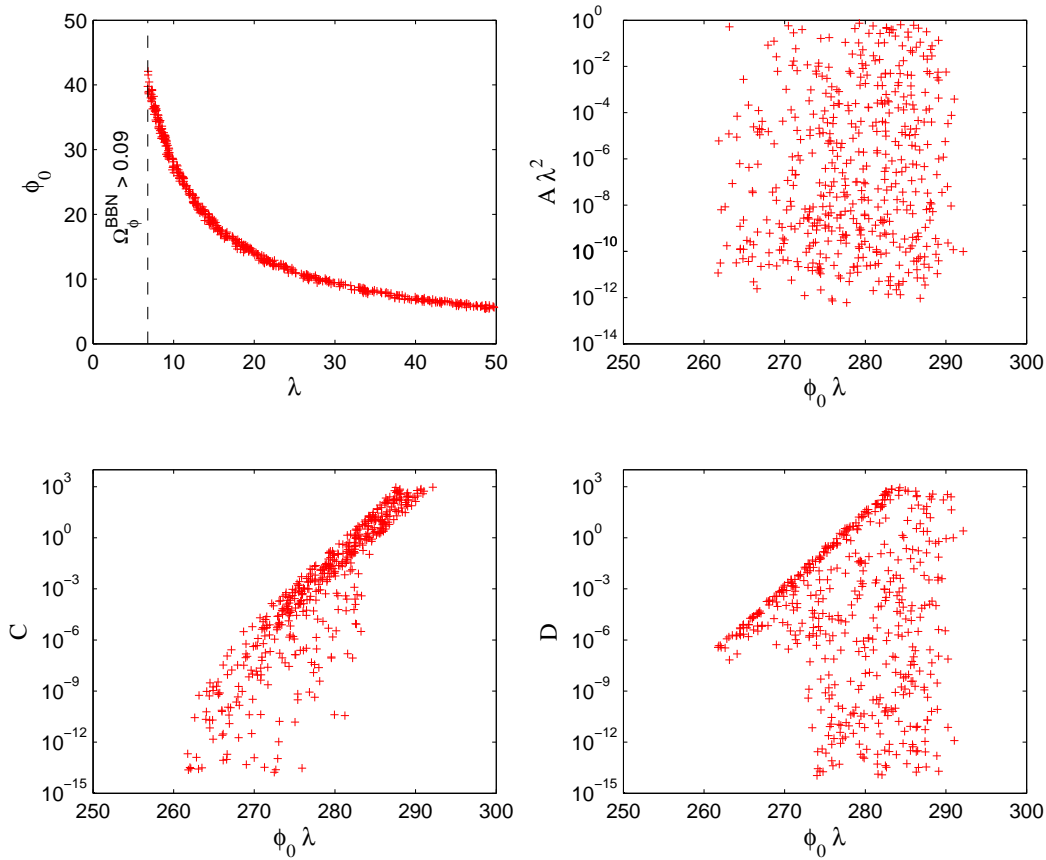}
\caption{Parameter space consistent with all the observational bounds considered, for the AS2 model.}
 \label{fig:AS22}
\end{figure*}

Our results for the AS2 model are shown in Figures \ref{fig:AS21} and \ref{fig:AS22}. For the sake of comparison with the AS1 model, we use the same combination of parameters $A \lambda^2$ and $\phi_0 \lambda$. In this case, we find
\begin{align}\label{eq:boundperm2}
6.8 &\lesssim \,\lambda\, \lesssim 50.1\,,\nonumber\\
261.7 &\lesssim\, \phi_0 \lambda \lesssim 292.2\,,\nonumber\\
6.1 \times 10^{-13} &\lesssim A \lambda^2 \lesssim 7.4 \times 10^{-2}\,,\\
1.7 \times 10^{-14} &\lesssim\, C \lesssim 9.4 \times 10^{2}\,,\nonumber\\
1.1 \times 10^{-14} &\lesssim\, D \lesssim 8.7 \times 10^{2}\,.\nonumber
\end{align}

\begin{figure*}[t]
\includegraphics[width=16cm]{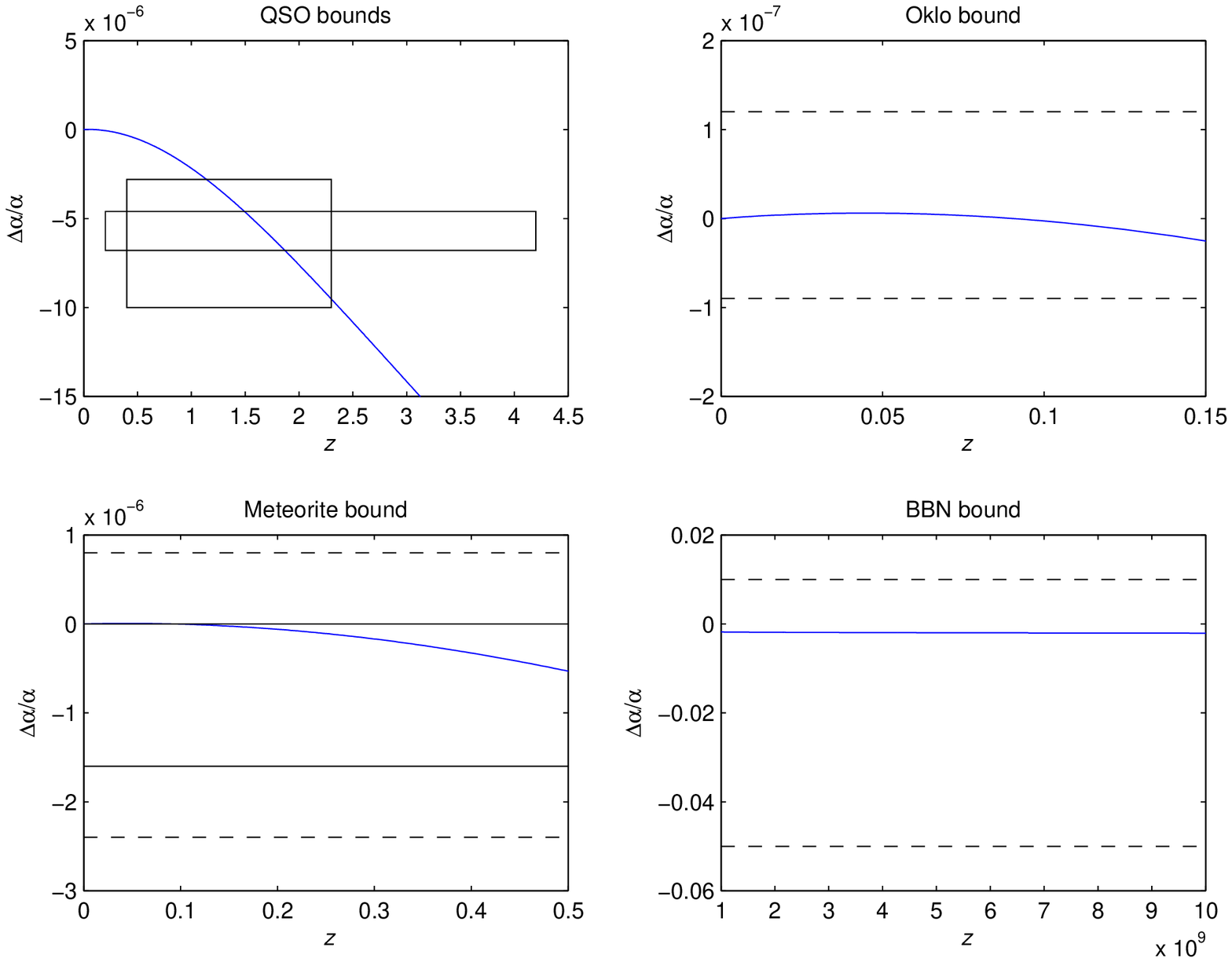}
\caption{Evolution of $\Delta \alpha/\alpha$ for $\xi=2.0\times 10^{-4}$ and the particular set of parameters $\lambda=6.8, \phi_0\lambda=272.5$ and $A \lambda^2=0.986$, in the transient acceleration case of the AS1 model. The boxes and horizontal lines correspond to the observational bounds given in Eqs.~(\ref{murphy})-(\ref{BBN}).}
\label{fig:AS1e}
\end{figure*}

Finally, in Figure \ref{fig:AS1e} we give an example of how, for particular values of the potential parameters, the AS1 model fits the observational bounds on the variation of the fine structure constant. 

\section{Conclusion}

The time variation of fundamental constants is a common prediction of theories that attempt to unify the four fundamental interactions. The experimental bounds obtained on the variation of these constants are therefore a useful tool for testing the validity of these theories. In this work we have studied the implications of the coupling of electromagnetism to quintessence fields. In our analysis we have considered two simple quintessence models with a modified exponential potential~\cite{skordis1,skordis2}, with the aim to constrain the parameter space using the observational data on the variation of the fine structure constant $\alpha$ in combination with the 5-year data from WMAP. Our results are summarized in Eqs.~(\ref{eq:boundperm1})-(\ref{eq:boundperm2}).

We should point out that, in constraining the models, the observational data on the variation of $\alpha$, obtained from the quasar absorption systems [cf. Eqs.(\ref{murphy}) and (\ref{chand})], turned out to be crucial when establishing limits on the parameters of the AS1 and AS2 models. This is due to the fact that the QSO bounds imply a non-vanishing value of $\Delta\alpha/\alpha$ at redshift $0.2 \lesssim z \lesssim 4.2$, which in turn requires a non-vanishing coupling $\xi$ between the scalar and electromagnetic fields. On the other hand, it was not possible to use in our approach the Oklo, meteorite, BBN and CMB bounds to put any tight constraint on the model parameters, since these observational bounds are all consistent with no variation of $\alpha$ at their corresponding redshift scales. 
\vspace{0.5cm}
\begin{acknowledgments}

The work of M.C.B. is partially supported by the FCT project POCI/FIS/56093/2004.

\end{acknowledgments}


\end{document}